# DECENTRALIZED SUBSPACE PURSUIT FOR JOINT SPARSITY PATTERN RECOVERY

*Gang Li[1], Thakshila Wimalajeewa[2], Pramod K. Varshney[2]*

1. Tsinghua University, Beijing 100084, China
2. Syracuse University, Syracuse, NY 13244, USA

**ABSTRACT**

To solve the problem of joint sparsity pattern recovery in a decentralized network, we propose an algorithm named decentralized and collaborative subspace pursuit (DCSP). The basic idea of DCSP is to embed collaboration among nodes and fusion strategy into each iteration of the standard subspace pursuit (SP) algorithm. In DCSP, each node collaborates with several of its neighbors by sharing high-dimensional coefficient estimates and communicates with other remote nodes by exchanging low-dimensional support set estimates. Experimental evaluations show that, compared with several existing algorithms for sparsity pattern recovery, DCSP produces satisfactory results in terms of accuracy of sparsity pattern recovery with much less communication cost.

## 1. INTRODUCTION

Compressive sensing (CS) refers to the idea that a sparse signal can be accurately recovered from a small number of measurements [1]-[3]. It has been shown that CS is potentially useful in a wide range of applications including medical imaging [4], radar imaging [5], source localization [6], and spectrum sensing [7]. In particular, CS provides a new perspective for data reduction in sensor network applications without compromising performance [8]-[10].

Consider a network composed of $L$ distributed nodes. The measurements collected at the $l$-th node are given by

$$\mathbf{y}_l = \mathbf{A}_l \mathbf{x}_l \qquad (1)$$

where $\mathbf{y}_l$ is an $M \times 1$ measurement vector, $\mathbf{A}_l$ is an $M \times N$ dictionary matrix, $\mathbf{x}_l$ is an $N \times 1$ vector which has $K$ nonzero entries with indices listed in a support set $\mathbf{S}$. Assume that all $\{\mathbf{x}_l, l=1, 2, \cdots, L\}$ have the same sparsity pattern, i.e., $\mathbf{S}=\{i: \mathbf{x}_l(i) \neq 0$ for $i=1, 2, \cdots, N\}$ with cardinality $|\mathbf{S}|=K$. Our goal is to recover the support set $\mathbf{S}$ using $\{\mathbf{y}_l, \mathbf{A}_l, l=1, 2, \cdots, L\}$ in the case that $N>M \geqslant 2K$. Once $\mathbf{S}$ is determined, the nonzero entries in $\mathbf{x}_l$ can be recovered by least squares estimation.

In centralized processing, it is required for each node to transmit its measurement vector and dictionary matrix to a central fusion center. Due to practical constraints on communication bandwidth and computational capacity, recovering sparsity pattern in a decentralized manner is more efficient. The joint orthogonal matching pursuit (JOMP) and the joint subspace pursuit (JSP) algorithms were presented in [11], where each node estimates the support set independently, and then the global estimate of the support set was obtained by fusing all individual estimates by majority rule. In [12], the distributed version of JSP, the distributed subspace pursuit (DiSP) algorithm, was developed by restricting the number of connectable neighbors for each node. DiSP reduces the requirement of network connectivity, however, at the cost of decreased accuracy of sparsity pattern recovery. The algorithms in [13]-[17] achieve better accuracy in sparsity pattern recovery compared to that in [11]-[12], by embedding collaboration among nodes into the iterative solution process at each node. In the distributed basis pursuit algorithms [13]-[14], each node solves for a local sparse solution by convex optimization and refines the solution by communicating with other nodes in an iterative manner. These convex optimization based algorithms require large computational resources. In the simultaneous orthogonal matching pursuit (SOMP) [15]-[16] and the simultaneous subspace pursuit (SSP) algorithms [17], collaboration among nodes at each iteration is also employed, but local estimates of the support set are obtained by greedy pursuit procedures. Generally speaking, SOMP and SSP are computationally much simpler than the convex optimization based algorithms in [13]-[14].

To implement the algorithms in [13]-[17] in a decentralized network, at each iteration, each node has to send $O(N)$-length vectors consisting of the locally estimated coefficients to all the other nodes in the network. Before the iterative process converges, these $O(N)$-length messages are probably not sparse, and therefore, the total number of messages to be transmitted is considerably large. Thus, applying the algorithms in [13]-[17] to a decentralized network requires large communication bandwidth. A communication efficient algorithm, the distributed and collaborative orthogonal matching pursuit (DCOMP) algorithm, was proposed in [18], where the reduction of communication cost comes from the fact that transmission of $O(N)$-length-vectors is restricted to a small neighborhood surrounding each node. However, due to lack of backtracking operations, once an index is considered as reliable and selected by DCOMP, it is not removed from the support set. This means that the strategy of index selection in DCOMP is too restrictive, and therefore, a larger number of measurements is required to guarantee the success of sparsity pattern recovery.

In this paper, we develop a new communication-efficient algorithm named decentralized and collaborative subspace pursuit (DCSP) for joint sparsity pattern recovery. The way that nodes in DCSP collaborate is different from that in JOMP and JSP. Similar to the algorithms in [13]-[17], DCSP also embeds the collaboration among nodes into the iterative solution process and, therefore, is superior to JOMP and JSP in terms of accuracy of sparsity pattern

The work of Dr. Li was supported in part by National Natural Science Foundation of China under Grant 41271011, and in part by 973 Program under Grant 2010CB731901, and in part by Program for New Century Excellent Talents in University under Grant NCET-11-0270, and in part by Tsinghua University Initiative Scientific Research Program. The work of Dr. Wimalajeewa and Dr. Varshney was supported in part by National Science Foundation Award No. 1307775.

recovery. Different from the unrestricted collaboration among nodes in SOMP [15]-[16] and SSP [17], at each iteration of DCSP, each node shares O($N$)-length messages only with a few of its neighboring nodes and communicates with other remote nodes by exchanging local $K$-length estimates of the support set. Therefore, compared to SOMP and SSP, DCSP significantly reduces the number of messages to be transmitted and, accordingly, requires less communication bandwidth of the network. Different from DCOMP in which all of the index estimates from past iterations are deemed reliable, DCSP is capable of removing wrong index estimates during each iteration by reevaluating the reliability of the previously estimated support set and fusing all the local index estimates. Therefore, compared to DCOMP, DCSP provides much better accuracy of sparsity pattern recovery at a comparable communication cost.

The rest of this paper is organized as follows. In Section 2, the implementation of SSP in a decentralized network is introduced. In Section 3, the DCSP algorithm is proposed for joint sparsity pattern recovery. Simulation results are provided in Section 4 and concluding remarks are given in Section 5.

## 2. SSP IN A DECENTRALIZED NETWORK

The original SSP algorithm was proposed in [17] for centralized processing. In this section, we consider the implementation of SSP in a decentralized network consisting of $L$ nodes. We assume that there is no fusion center and the sparsity pattern estimation has to be performed via collaboration among nodes. To achieve the same accuracy of sparsity pattern recovery as with the centralized SSP algorithm, all nodes in the decentralized network need to exchange local processing results with each other at each iteration. The operations at the $l$-th node are summarized in Algorithm 1. We do not consider any specific communication protocol here and assume the worst-case scenario in terms of communication complexity that all the nodes communicate with each other one-by-one. To simplify the presentation, we define the following notations.

- proj($\mathbf{y}, \mathbf{A}$) = $[\mathbf{A}^H\mathbf{A}]^{-1}\mathbf{A}^H\mathbf{y}$ calculates projection coefficients of a vector $\mathbf{y}$ onto the column space of the matrix $\mathbf{A}$. $(\cdot)^H$ denotes conjugate transpose.
- resid($\mathbf{y}, \mathbf{A}$) = $\mathbf{y} - \mathbf{A}[\mathbf{A}^H\mathbf{A}]^{-1}\mathbf{A}^H\mathbf{y}$ outputs the projection residual vector.
- max_ind($\mathbf{y}, K$)={$K$ indices corresponding to the largest magnitude entries in the vector $\mathbf{y}$}.
- max_occ($\mathbf{S}, K$) ={$K$ elements that have the highest frequency of occurrence in the set $\mathbf{S}$}.
- $\mathbf{A}(\mathbf{S})$ denotes a sub-matrix composed of the columns of $\mathbf{A}$ indexed by the set $\mathbf{S}$.
- $\mathbf{y}(\mathbf{S})$ denotes a sub-vector composed of the entries of $\mathbf{y}$ indexed by the set $\mathbf{S}$.
- $\mathbf{G}$={1, 2, ···, $L$} records the indices of all nodes.
- superscript $t$ denotes the iteration counter.

Our focus is on the communication cost during the collaboration among nodes. The communication cost can be considered proportional to the number of messages to be transmitted. The communication between nodes in the network appear in Steps 1), 3), 5) and 7) of SSP, and the lengths of the messages transmitted from each node are $N$, $N$, $2K$ and 1, respectively. Thus, the total number of messages transmitted from all nodes is

$$C_{SSP} = \left[N + T_{SSP}(N + 2K + 1)\right](L-1)L , \qquad (2)$$

**Algorithm 1 The SSP algorithm at the $l$-th node**

Input: $K$, $\mathbf{y}_l$, $\mathbf{A}_l$.
Initialization:
1) Send the vector $\mathbf{c}_l^0 = |\mathbf{A}_l^H\mathbf{y}_l|$ to and receive $\mathbf{c}_j^0$ from the $j$-th node, for all $j \in \mathbf{G}\backslash\{l\}$.
2) Let $\mathbf{S}^0$ = max_ind($\sum_{l \in \mathbf{G}} \mathbf{c}_l^0$, $K$); calculate the residual $\mathbf{r}_l^0$ = resid($\mathbf{y}_l$, $\mathbf{A}_l(\mathbf{S}^0)$).

Iteration: at the $t$-th iteration ($t \geq 1$)

3) Send the vector $\mathbf{c}_l^t = |\mathbf{A}_l^H\mathbf{r}_l^{t-1}|$ to and receive $\mathbf{c}_j^t$ from the $j$-th node, for all $j \in \mathbf{G}\backslash\{l\}$.
4) Let $\tilde{\mathbf{S}}^t = \mathbf{S}^{t-1} \cup$ max_ind($\sum_{l \in \mathbf{G}} \mathbf{c}_l^t$, $K$).
5) Send the vector $\mathbf{d}_l^t = \text{proj}(\mathbf{y}_l, \mathbf{A}_l(\tilde{\mathbf{S}}^t))$ to and receive $\mathbf{d}_j^t$ from the $j$-th node, for all $j \in \mathbf{G}\backslash\{l\}$.
6) Let $\mathbf{S}^t$ = max_ind($\sum_{l \in \mathbf{G}} |\mathbf{d}_l^t|$, $K$); update the residual $\mathbf{r}_l^t$ = resid($\mathbf{y}_l$, $\mathbf{A}_l(\mathbf{S}^t)$).
7) Send the value of $\|\mathbf{r}_l^t\|_2^2$ to and receive $\|\mathbf{r}_j^t\|_2^2$ from the $j$-th node, for all $j \in \mathbf{G}\backslash\{l\}$.
8) If $\sum_{l \in \mathbf{G}} \|\mathbf{r}_l^t\|_2^2 \geq \sum_{l \in \mathbf{G}} \|\mathbf{r}_l^{t-1}\|_2^2$, let $\mathbf{S}^t = \mathbf{S}^{t-1}$ and stop; otherwise, let $t = t+1$, and return to Step 3).

Output: The estimated support set $\mathbf{S}^t$.

where $T_{SSP}$ is the number of iterations needed to successfully recover the support set by SSP and its value depends on the amplitude distribution of the sparse signal [19]. The total communication cost in (2) will dramatically increase as the network connectivity increases.

## 3. THE DCSP ALGORITHM

Considering the tradeoff between the communication cost and the recovery accuracy, in this section we present the DCSP algorithm, which provides satisfactory accuracy of sparsity pattern recovery and requires much less communication overhead compared to SSP. Note that the communication cost in SSP is dominated by the transmissions of $N$-length messages among nodes. From this observation, it is desirable to reduce the communication cost by restricting exchanges of $N$-length messages among nodes. In the DCSP algorithm, each node is allowed to share O($N$)-length messages only with a few of its neighboring nodes and communicates with other remote nodes in the network by transmitting and receiving $K$-length messages. The indices of the neighbors of the $l$-th node are recorded in the set $\mathbf{G}_l$, which also contains the $l$-th node itself. When $\mathbf{G}_l = \mathbf{G}$ for all $l \in \mathbf{G}$, the DCSP algorithm will degenerate into the decentralized version of SSP described in Section 2.

In DCSP summarized in Algorithm 2, each node first collaborates with a few of its neighbors to obtain the local estimate of the support set and then fuses all local estimates received from all nodes in the network by majority role, as described below. At each iteration, each node communicates with its neighbors twice and broadcasts to all nodes twice. After initialization, the first collaboration among neighboring nodes appears in Step 5) of DCSP, where the $l$-th node shares an $N$-length correlation coefficient vec-

**Algorithm 2 The DCSP algorithm at the $l$-th node**

Input: $K$, $\mathbf{y}_l$, $\mathbf{A}_l$.
Initialization:
1) Send the vector $\mathbf{c}_l^0 = |\mathbf{A}_l^H \mathbf{y}_l|$ to and receive $\mathbf{c}_j^0$ from the $j$-th node, for all $j \in \mathbf{G}_l \backslash \{l\}$.
2) Let $\mathbf{\Gamma}_l^0 = \text{max\_ind}(\sum_{l \in \mathbf{G}_l} \mathbf{c}_l^0, K)$.
3) Send $\mathbf{\Gamma}_l^0$ to and receive $\mathbf{\Gamma}_j^0$ from the $j$-th node, for all $j \in \mathbf{G} \backslash \{l\}$.
4) Let $\mathbf{\Gamma}^0 = \{\mathbf{\Gamma}_1^0, \mathbf{\Gamma}_2^0, \cdots, \mathbf{\Gamma}_L^0\}$ and $\mathbf{S}^0 = \text{max\_occ}(\mathbf{\Gamma}^0, K)$; set the residual $\mathbf{r}_l^0 = \text{resid}(\mathbf{y}_l, \mathbf{A}_l(\mathbf{S}^0))$.

Iteration: at the $t$-th iteration ($t \geq 1$)
5) Send the vector $\mathbf{c}_l^t = |\mathbf{A}_l^H \mathbf{r}_l^{t-1}|$ to and receive $\mathbf{c}_j^t$ from the $j$-th node, for all $j \in \mathbf{G}_l \backslash \{l\}$.
6) Let $\tilde{\mathbf{S}}_l^t = \mathbf{S}^{t-1} \cup \text{max\_ind}(\sum_{l \in \mathbf{G}_l} \mathbf{c}_l^t, K)$; calculate the projection coefficients $\overline{\mathbf{x}}_l^t = \text{proj}(\mathbf{y}_l, \mathbf{A}_l(\tilde{\mathbf{S}}_l^t))$.
7) Send $\overline{\mathbf{x}}_l^t$ to and receive $\overline{\mathbf{x}}_j^t$ from the $j$-th node, for all $j \in \mathbf{G}_l \backslash \{l\}$.
8) Let $\mathbf{\Gamma}_l^t = \text{max\_ind}(\sum_{l \in \mathbf{G}_l} |\overline{\mathbf{x}}_l^t|, K)$.
9) Send $\mathbf{\Gamma}_l^t$ to and receive $\mathbf{\Gamma}_j^t$ from the $j$-th node, for all $j \in \mathbf{G} \backslash \{l\}$.
10) Let $\mathbf{\Gamma}^t = \{\mathbf{\Gamma}_1^t, \mathbf{\Gamma}_2^t, \cdots, \mathbf{\Gamma}_L^t\}$ and $\mathbf{S}^t = \text{max\_occ}(\mathbf{\Gamma}^t, K)$; update the residual $\mathbf{r}_l^t = \text{resid}(\mathbf{y}_l, \mathbf{A}_l(\mathbf{S}^t))$.
11) Send $\|\mathbf{r}_l^t\|_2^2$ to and receive the $\|\mathbf{r}_j^t\|_2^2$ from the $j$-th node, for all $j \in \mathbf{G} \backslash \{l\}$.
12) If $\sum_{l \in \mathbf{G}} \|\mathbf{r}_l^t\|_2^2 \geq \sum_{l \in \mathbf{G}} \|\mathbf{r}_l^{t-1}\|_2^2$, let $\mathbf{S}^t = \mathbf{S}^{t-1}$ and stop; otherwise, let $t = t+1$, and return to Step 5).

Output: The estimated support set $\mathbf{S}^t$.

Table 1 Comparison in terms of communication cost

| Algorithm | Number of transmitted messages |
|---|---|
| JSP&JOMP | $K(L-1)L$ |
| SOMP | $KN(L-1)L$ |
| DCOMP | $T_{DCOMP}((g-1)NL + (L-1)L)$ |
| SSP | $[N + T_{SSP}(N+2K+1)](L-1)L$ |
| DCSP | $(g-1)NL + K(L-1)L$ <br> $+ T_{DCSP}L((g-1)(N+2K) + (K+1)(L-1))$ |

tor $\mathbf{c}_l^t$ with its neighbors. The coefficients in $\mathbf{c}_l^t$ are probably not sparse before the iterations converge, so transmission of the entire $\mathbf{c}_l^t$ is necessary. The neighboring nodes collaborate once again in Step 7) of DCSP, where the $l$-th node shares a $2K$-length projection coefficient vector $\overline{\mathbf{x}}_l^t$ with its neighbors. By such two-stage collaboration, the $l$-th node selects $K$ indices from $\tilde{\mathbf{S}}_l^t$ as the local estimate of the support set by finding a $K$-dimensional subspace that the local measurements lie in. Then in Step 9) of DCSP, the $l$-th node broadcasts the $K$-length vector $\mathbf{\Gamma}_l^t$ (i.e. the local estimate of the support set) to all nodes in the network. As a result, the index set $\mathbf{\Gamma}^t$ in Step 10) of DCSP is the same for all the nodes in the network. Fusion of all local estimates of the support set is performed according to majority rule. In Step 10) of DCSP, $K$ indices that have the highest frequency of occurrence are recorded in the set $\mathbf{S}^t$. Thus, $\mathbf{S}^t$ represents the global estimate of the support set. In Step 11) of DCSP, each node broadcasts once again to report the local recovery error. When the global recovery error reaches the minimum, iterations at all the nodes are terminated. From all the steps of DCSP, the total number of transmitted messages can be calculated as

$$C_{DCSP} = N\sum_{l \in \mathbf{G}}(g_l - 1) + K(L-1)L \\ + T_{DCSP}\left((N+2K)\sum_{l \in \mathbf{G}}(g_l - 1) + (K+1)(L-1)L\right), \quad (3)$$

where $T_{DCSP}$ is the number of iterations needed to successfully recover the support set by DCSP.

In what follows, we compare different algorithms in terms of the number of messages to be transmitted. Without loss of generality, a symmetric network is considered and two assumptions are made: 1) $g_l = g$ for all $l \in \mathbf{G}$, and 2) the neighbors of the $l$-th node are indexed by $\{\text{mod}(l+1,L)+1, \text{mod}(l+2,L)+1, \cdots, \text{mod}(l+g-1,L)+1\}$, where $\text{mod}(\cdot)$ is the modulus operation. The numbers of messages to be transmitted in different algorithms are listed in Table 1, where $T_{DCOMP}$ is the number of iterations needed for successful sparsity pattern recovery by DCOMP. JOMP and JSP have the smallest number of transmitted messages in Table 1, since in these two algorithms, support set fusion is performed after each node independently completes the local estimation of the support set. However, as shown in the next section, the recovery accuracies of JOMP and JSP are unsatisfactory. Since $K \ll N$, $g < L$ and all values of $T_{DCOMP}$, $T_{SSP}$ and $T_{DCSP}$ are in the order of $O(K)$, we can see that the communication cost of DCSP is much less than that of SOMP and SSP and in the same order of magnitude as that of DCOMP.

It is also worth discussing the recovery accuracies of DCOMP and DCSP, since their communication costs are comparable. In contrast to DCOMP, DCSP is capable of removing poor index estimates from the support set estimate. This is achieved by two operations: 1) in Steps 6) of DCSP, each node reevaluates the reliability of previous index estimates and newly added indices; and 2) in Step 10) of DCSP, poor local estimates of indices corresponding to low frequency of occurrence are rejected by majority voting. Thus, the recovery accuracy of DCSP is expected to be better than that of DCOMP, which will also be shown by simulations in the next section.

## 4. SIMULATION RESULTS

In the first experiment, we evaluate the sparsity pattern recovery capability of different algorithms. Consider a network composed of 6 nodes, i.e., $L=6$. With the fixed sparsity $K=10$ and the length of sparse solution at each node $N=200$, simulations are carried out as follows.
a) Choose a value of $M$ such that $M \geq 2K$.
b) Randomly generate a set of $M \times N$ dictionary matrices $\{\mathbf{A}_l, l=1, 2, \cdots, L\}$ from the standard independent and identically distributed (iid) Gaussian ensemble.
c) Randomly select $K$ indices from $\{1, 2, \cdots, N\}$ as the support set $\mathbf{S}$, and draw the entries of $\mathbf{x}_l$ supported on $\mathbf{S}$ from the standard

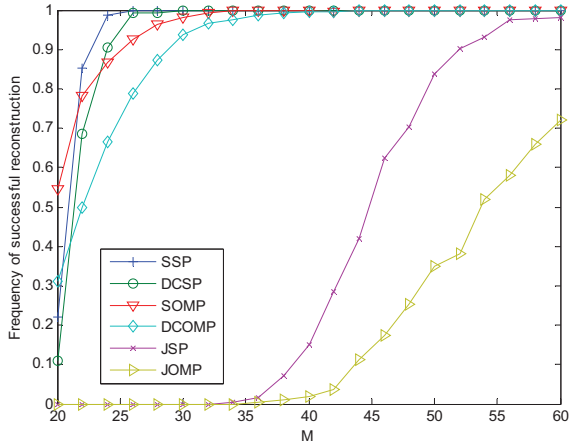

Fig. 1 Frequency of successful reconstruction vs number of measurements per node.

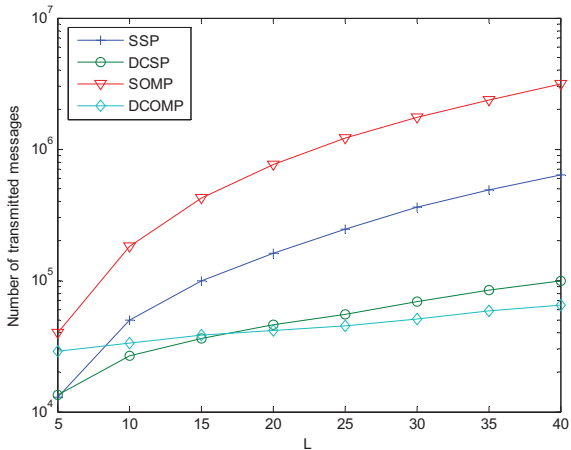

Fig. 2 Number of messages to be transmitted vs network scale

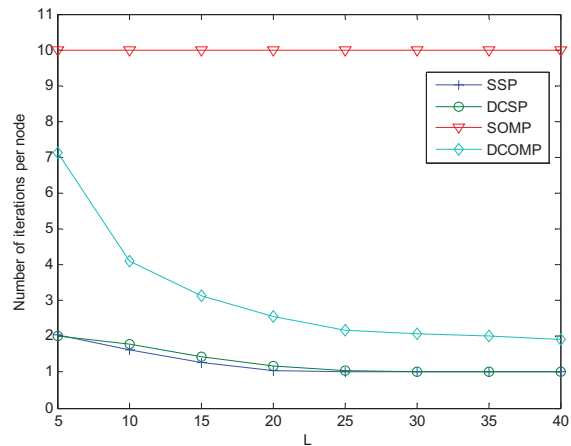

Fig. 3 Number of required iterations vs network scale.

iid Gaussian ensemble, for $l=1, 2, \cdots, L$.

d) Generate the measurement vectors $\mathbf{y}_l = \mathbf{A}_l \mathbf{x}_l$, $l=1, 2, \cdots, L$. Apply different algorithms to recover the support set. If the support set estimate is equal to $\mathbf{S}$, the recovery is considered as successful.

e) Repeat the Steps b) to d) 500 times for each value of $M$, and count the frequency of successful reconstruction.

In Fig. 1, we compare DCSP with five algorithms for sparsity pattern recovery (i.e. SSP [17], SOMP [16], DCOMP [18], JOMP and JSP [11]) in terms of accuracy of sparsity pattern recovery. Assuming that all of the 6 nodes are fully collaborative, SSP requires $M \geq 26$ to achieve success rate greater than 99.8%, which is the best performance in Fig. 1. Under the same assumption, SOMP requires $M \geq 34$ to achieve success rate greater than 99.8%. In both DCSP and DCOMP, the number of neighbors is set to $g=3$ for each node. DCSP and DCOMP require $M \geq 30$ and $M \geq 44$ to achieve success rates greater than 99.8%, respectively, which means that DCSP has much better sparsity pattern recovery capability than DCOMP. It is worth emphasizing that, due to the capability of removing wrong index estimates from the support set, DCSP based on full collaboration among 3 nodes yields slightly higher success rate than SOMP based on full collaboration among all 6 nodes.

In the second experiment, we compare communication costs and convergence speeds of four algorithms (i.e. SSP, SOMP, DCOMP and DCSP) that have higher success rates in Fig. 1, i.e., JOMP and JSP are not considered. Let $M=50$, $N=200$ and $K=10$. Assume that the network scale is increasing, i.e., $L$ is varying from 5 to 40, and the number of neighbors is fixed at $g=3$ for each node. The number of messages to be transmitted and the required number of iterations for the four algorithms are plotted in Fig. 2 and Fig. 3, respectively. The simulation implementation is the same as that in the previous experiment, and every point is obtained by averaging over 100 trials. Fig. 2 demonstrates the superiority of DCSP to SOMP and SSP in terms of communication efficiency. The reason is that DCSP replaces a large number of exchanges of $N$-length messages among nodes with exchanges of $K$-length messages. The communication cost of DCSP is comparable with that of DCOMP: in a small scale network, the number of transmitted messages of DCSP is slightly smaller than that of DCOMP, while the situation is reversed in a large scale network. Fig 3 plots the number of iterations required for different algorithms versus the number of nodes. Thanks to the effective collaboration among nodes, SSP and DCSP have faster convergence speeds compared to DCOMP and SOMP, and they are even capable of successfully recovering the support set from only one iteration when the number of nodes is large enough.

## 5. CONCLUSION

In this paper, we focused on the recovery of sparsity pattern in a decentralized network. By embedding collaboration among neighboring nodes and fusion strategy into each iteration of the standard SP algorithm, we developed an algorithm named DCSP for decentralized estimation of the sparsity pattern. In the DCSP algorithm, each node collaborates with a few of its neighbors by sharing $O(N)$-length messages and communicates with other remote nodes by exchanging $K$-length messages. Simulation results show that, compared with other similar algorithms, DCSP provides satisfactory recovery accuracy and has less communication cost. Future work includes convergence analysis, study of robustness to quantization errors, and extension to more complicated sparse structures. Our approach presented here can be easily combined with the compressive sampling matching pursuit (CoSaMP) algorithm [20], since CoSaMP and SP are very similar to each other.